\documentclass[a4paper,12pt]{article}

\usepackage{graphicx}
\usepackage{array}
\usepackage{amsmath}
\usepackage{amssymb}
\usepackage{latexsym}
\usepackage{subfigure}
\usepackage{color}

\makeatletter
\@addtoreset{equation}{section}
\makeatother

\date{empty}
\begin{document}
\begin{titlepage}
\null
\begin{flushright}
TIT/HEP-638 \\
Sep, 2014
\end{flushright}
\vskip 0.5cm
\begin{center}
{\Large \bf
Supersymmetry and R-symmetry Breaking in \\
\vskip 0.3cm
Meta-stable Vacua at Finite Temperature and Density
}
\vskip 1.1cm
\normalsize
\renewcommand\thefootnote{\alph{footnote}}

{\large
Masato Arai$^{\dagger}$\footnote{masato.arai(at)fukushima-nct.ac.jp},
Yoshishige Kobayashi$^\ddagger $\footnote{yosh(at)th.phys.titech.ac.jp}
and Shin Sasaki$^\sharp$\footnote{shin-s(at)kitasato-u.ac.jp}
}
\vskip 0.7cm
  {\it
  $^\dagger$
  Fukushima National College of Technology \\ 
  \vskip -0.2cm
  Iwaki, Fukushima 970-8034, Japan \\
  \vskip 0.1cm
  $^\ddagger$
  Department of Physics, Tokyo Institute of Technology \\
  \vskip -0.2cm
  Tokyo 152-8551, Japan \\
  \vskip 0.1cm 
  $^\sharp$
  Department of Physics,  Kitasato University \\
  \vskip -0.2cm
  Sagamihara 252-0373, Japan
}
\vskip 0.5cm
\begin{abstract}
\begin{sloppypar}
We study a meta-stable supersymmetry-breaking vacuum in a
generalized O'Raifeartaigh model 
at finite temperature and chemical potentials.
Fields in the generalized O'Raifeartaigh model possess different R-charges
to realize R-symmetry breaking. Accordingly, at finite density and temperature, 
the chemical potentials have to be introduced in a non-uniform way.
Based on the formulation elaborated in
our previous work we study the one-loop thermal effective potential including 
the chemical potentials in the generalized O'Raifeartaigh model.
We perform a numerical analysis and find that 
the R-symmetry breaking vacua,
which exist at zero temperature and zero chemical potential,
are destabilized for some parameter regions.
In addition, we find that there are parameter regions where
new R-symmetry breaking vacua are realized even at high temperature
by the finite density effects.
\end{sloppypar}
\end{abstract}
\end{center}

\end{titlepage}

\newpage
\setcounter{footnote}{0}
\renewcommand\thefootnote{\arabic{footnote}}
\pagenumbering{arabic}
\section{Introduction}
Dynamical supersymmetry breaking is an important topic in
particle physics.
For phenomenological model building, 
recent studies reveal that  supersymmetry breaking at a global vacuum is not
necessary instead, supersymmetry breaking at a meta-stable local vacuum is sufficient
\cite{Intriligator:2006dd}.
Even though models admit supersymmetric vacua (true vacua), 
if a meta-stable vacuum (a false vacuum) is long-lived,
the models are phenomenologically viable.
Meta-stable supersymmetry breaking vacua make model building easier and 
many phenomenological models that admit such vacua are proposed.
In supersymmetric model building, in addition to supersymmetry,
the R-symmetry also must be broken 
because it is the necessary condition for 
non-zero Majorana gaugino masses.
However, it is widely known that the R-symmetry is not broken 
in general for spontaneous F-term supersymmetry breaking models \cite{Nelson:1993nf}.
In order to overcome this problem, a model which is a generalization of the
O'Raifeartaigh model has been proposed \cite{Shih:2007av}.
This generalized O'Raifeartaigh model consists of some fields with R-charges
other than 0 or 2 and one (pseudo) modulus field $X$ with the R-charge 2.
The model breaks supersymmetry spontaneously 
and the modulus field $X$ parametrizes a continuous space of degenerate 
vacua with non-zero tree-level energy.
There are also supersymmetric vacua in runaway directions. 
Taking one-loop corrections into account,
the degeneracy of vacua is resolved and a local minimum appears at a non-zero
vacuum expectation value (VEV) of $X$ by which the R-symmetry is spontaneously broken.
With a suitable choice of parameters in the model, it is shown that the local 
minimum can be long-lived and therefore it is a meta-stable vacuum. 
Further generalization of the O'Raifeartaigh model with global symmetries is also 
studied \cite{Fe}.

It is considered that supersymmetry holds in the early Universe and it is 
spontaneously broken by some mechanism in the evolution of the Universe.
In order to study breaking of supersymmetry in the expansion of the Universe, 
it is necessary to take temperature and density into account, since they may give 
significant effects on the phase structure of supersymmetry
and R-symmetry breaking vacua.
The thermal effective potential of the generalized O'Raifeartaigh models 
for vanishing chemical potentials has been studied 
{\cite{AbChJaKh1,CrFoWa,FiKaKrMaTo,AbJaKh,MoSc,Ka,KaPa,ArKoSa}.
The thermal history of the meta-stable vacua is examined
and the possibility of the first order phase transition in a certain parameter
region is discussed.
It is shown there that the R-symmetry of the meta-stable vacuum is generically 
restored at high temperature.
For the case with finite chemical potentials, the thermal history of the effective 
potential of a supersymmetric model with a single flavor is studied \cite{RiSe}.
This model shows that there exists a vacuum where
the R-symmetry is broken even at high temperature.
However, spontaneous supersymmetry breaking is not considered though the effects of
the soft supersymmetry breaking are mentioned.
It is interesting to investigate how the chemical potential affects 
the R-symmetry breaking in spontaneous supersymmetric breaking models such as the
generalized O'Raifeartaigh models.
However, in such models each field has different
values of chemical potentials which are proportional to the global R-charges.
We call this {\it non-uniform chemical potential}.
It makes it difficult to derive the thermal effective potential with the non-uniform
chemical potentials.
This is because the mass matrices and the matrix associated with the
non-uniform chemical potential do not commute with each other 
detailed analysis of the sum over the Kaluza-Klein momentum 
appearing in the effective potential is needed. 
In \cite{Arai:2013ema} 
we have carefully performed summation of the Kaluza-Klein momentum and
have derived the general formula for the thermal effective
potential with non-uniform chemical potentials. 

The purpose of this paper is to study 
effects of finite temperature and non-uniform chemical
potentials on the meta-stable vacua in the generalized
O'Raifeartaigh model. We consider the simplest model proposed in 
\cite{Shih:2007av}.
Applying the general formula constructed in
\cite{Arai:2013ema}
we perform a numerical analysis of the thermal effective potential.
We will show that
a parameter region which allows the R-symmetry breaking reduces from one
in the case of vanishing chemical potentials studied in \cite{MoSc}.
Interestingly, there appears a new parameter region where 
the R-symmetry is broken even at high temperature.
This is a consequence of the non-uniform chemical potentials. 

The organization of this paper is as follows.
In the next section, we review the general formula for the thermal
effective potentials with the non-uniform chemical potentials.
In section 3, we show the vacuum structure of the model in the zero and
finite temperatures.
In section 4 we performed numerical calculations of the thermal
effective potential where the non-uniform chemical potentials are turned
on.
Section 5 is devoted to the conclusion and discussions. 

%
%
\section{Thermal effective potential with non-uniform chemical potentials}
In this section we briefly introduce the thermal effective potential
with the non-uniform chemical potentials elaborated in \cite{Arai:2013ema}.
We consider a model that consists of $N$ complex scalar fields $\phi_i \
(i=1, \cdots, N)$ and
$N$ Dirac fermionic fields $\psi_i \ (i=1, \cdots, N)$%
\footnote{
The number of the fermionic fields can be different from that of
the scalar fields in general. We have given the same number to both fields
for the later consideration of a supersymmetric theory.
}
with the Lagrangian
\begin{align}
\mathcal{L} = \partial_m \phi^{i} \partial^m
 \phi^{\dagger}_{i} 
- V(\phi, \phi^{\dagger}) 
+ i \bar{\psi}_{i} \gamma^m \partial_m \psi^{i} 
- \bar{\psi}_{i} \hat{M}^{i} {}_{j} (\phi, \phi^{\dagger})
 \psi^{j}. 
\end{align}
Here $V (\phi, \phi^{\dagger})$ and $\hat{M}^i {}_j (\phi,
\phi^{\dagger})$ are the scalar potential and the interaction matrix,
and $\gamma^m$ is the Dirac gamma matrix. 
The mass matrices $\hat{m}_B$ and $\hat{m}_F$ for the bosonic and fermionic fields are defined by 
\begin{align}
(\hat{m}^2_B)^i_{~j} 
= \left. \frac{\partial^2 V}{\partial
 \phi^{\dagger}_i \partial \phi^j} \right|_{\phi^i =
 \phi^i_{\mathrm{cl}}}, \qquad 
(\hat{m}_F)_{ij} = \left. \hat{M}^i {}_j \right|_{\phi^i = \phi^{i}_{\mathrm{cl}}},
\end{align}
where $\phi^i_{\mathrm{cl}}$ is the VEV of the scalar fields.
The action is invariant under the $U(1)$ global transformations $\phi^{i \, \prime} = e^{i
q^{i}} \phi^{i}$ and $\psi^{i \, \prime} = e^{i \tilde{q}^{i}}
\psi^{i}$. Here $q^i$ and $\tilde{q}^i$ are the $U(1)$ charges of the
corresponding fields.

The chemical potential is introduced by making the $U(1)$ symmetry be 
gauged \cite{Ac1,Ac2, HaLaMu}. The spacetime derivative in the kinetic term is replaced by the
gauge covariant derivative $D_m = \partial_m + i q A_m$ where $A_m$ is
the $U(1)$ gauge potential. 
The chemical potential $\mu$ is identified with the VEV of the zeroth
component of the gauge field $\langle A_m \rangle =
(i \mu, \vec{0})$. 
We note that since the $U(1)$ charge $q^i$ or $ \tilde{q}^i$ in each field is different in
general, the quadratic terms of bosonic and fermionic fields which depend on the chemical
potential are of the form
$ (\partial_0 \vec{\phi}^{\dagger}) \hat{\mu}_{B} \vec{\phi}
- \vec{\phi}^{\dagger} \hat{\mu}_{B}  (\partial_0 \vec{\phi} )
-\vec{\phi}^{\dagger} \hat{\mu}_{B}^2 \vec{\phi}$,
$\vec{\psi}^{\dagger} \hat{\mu}_{F} \vec{\psi}$ 
where $\hat{\mu}_B$ and $\hat{\mu}_F$ are diagonal matrices given by
\begin{align}
\hat{\mu}_B \equiv \mathrm{diag} (\mu_B^1, \quad \cdots, \mu_B^N),\quad 
\hat{\mu}_F \equiv \mathrm{diag} (\mu_F^1, \quad \cdots, \mu_F^N).
\end{align}
These matrices do
not commute with the mass matrices $\hat{m}_B^2$ and
$\hat{m}_F^2$ in general. 
We call $\hat{\mu}_B$ and $\hat{\mu}_F$
{\it non-uniform chemical potentials}. 
The thermal effective potential in the presence of the non-uniform
chemical potentials is studied in our previous paper \cite{Arai:2013ema}.
In order to introduce finite temperature $T$, we 
perform the Wick rotation $i x_0 = \tau$ and impose the (anti-)periodic conditions on the
fields $\phi_i (\tau, \vec{x}) = \phi_i (\tau + \beta, \vec{x})$,
$\psi_i (\tau, \vec{x}) = - \psi_i (\tau + \beta, \vec{x})$ where $\beta
= 1/T$. 
The one-loop part of the thermal effective potential is given by 
\begin{align}
V^{(1)} = V^{(1) \beta, \mu}_B + V^{(1) \beta, \mu}_F,
\end{align}
where $V^{(1) \beta, \mu}_B$ and $V^{(1) \beta, \mu}_F$ are
contributions which come from the bosonic and fermionic fields:
\begin{align}
& V^{(1) \beta, \mu}_B (\phi_{\mathrm{cl}}) = 
 \frac{1}{2\beta} 
\sum_{n= - \infty}^{\infty} \int \! \frac{d^3 p}{(2\pi)^3} \mathrm{Tr} 
\log (\omega_{B,n}^2 + \omega^2_{B,p}), 
\label{eq:boson_1-loop}
\\
& V^{(1) \beta, \mu}_F (\phi_{\mathrm{cl}}) = 
- \frac{2}{2\beta} 
\sum_{n= - \infty}^{\infty} \int \! \frac{d^3 p}{(2\pi)^3} \mathrm{Tr} 
\log (\omega_{F,n}^2 + \omega^2_{F,p}), 
\label{eq:fermion_1-loop}
\\
& \omega^2_{B,p} = p^2 \mathbf{1} + \hat{m}^2_B, \quad 
\omega^2_{F,p} = p^2 \mathbf{1} + \hat{m}^2_F, \quad \\
& \omega_{B,n} = 2 \pi \beta^{-1} n \mathbf{1} - i \hat{\mu}_{B}, \quad
\omega_{F,n} = 2 \pi \beta^{-1} n \mathbf{1} - i \hat{\mu}_{F},
\end{align}
where $\mathbf{1}$ is the $N \times N$ unit matrix.
This is the natural generalization of the 
thermal effective potential for single-flavor models \cite{Ac1, Ac2}.
Calculating the trace over the flavor indices
and the infinite sum over $n$ in the effective
potential is a little bit complicated since the matrices 
$\hat{m}_B, \hat{m}_F$
and $\hat{\mu}_B, \hat{\mu}_F$ do not commute with each other.
Taking this fact into account,
we have performed the calculation of the flavor trace and find the
following more explicit expressions,
\cite{Arai:2013ema}
\begin{align}
V^{(1) \beta, \mu} (\phi_{\mathrm{cl}}) =& \ V (\phi_{\mathrm{cl}}) + 
V_B^{(1) \, \beta, \mu} + V_F^{(1) \, \beta, \mu}, \\
V_B^{(1) \, \beta, \mu} (\phi_{\mathrm{cl}})
=& \ 
 \frac{1}{2 \beta} 
\int \! \frac{d^3 p}{(2 \pi)^3} 
\sum_{i = 1}^{2N} \log
\left(
\sin 
\left(
\frac{\beta \chi_{B,i}}{2}
\right)
\right), 
\label{eq:pot_bos}
\\
V_F^{(1) \, \beta, \mu} (\phi_{\mathrm{cl}}) =& \ 
\frac{2}{2 \beta} 
\int \! \frac{d^3 p}{(2 \pi)^3} 
\sum_{i = 1}^{2N} \log
\left(
\cos 
\left(
\frac{\beta \chi_{F,i}}{2}
\right)
\right).
\label{eq:pot_fer}
\end{align}
Here $\chi_{B,i}$ and $\chi_{F,i}$ are solutions to the following equations
respectively,
\begin{align}
\det (p^2 \mathbf{1} + \hat{m}_{B}^2 + (\chi_{B} \mathbf{1} - i \hat{\mu}_B)^2) = 0,\quad
\det (p^2 \mathbf{1} + \hat{m}_F^2 + (\chi_F \mathbf{1} - i \hat{\mu}_F)^2) = 0.
\label{eq:chi_eq}
\end{align}
Note that the chemical potentials should satisfy the bound such that
$\hat{m}_{B}^2 - \hat{\mu}_{B}^2$ 
and $\hat{m}_{F}^2 - \hat{\mu}_{F}^2$
are the positive definite matrices otherwise 
the effective potential becomes imaginary due to tachyonic modes\footnote{
Fermionic effective potential does not always become imaginary,
even if the bound $\hat{m}_{F}^2 - \hat{\mu}_{F}^2$ is not
satisfied. However this bound is a sufficient condition for the reality condition of the 
effective potential.
}.
We also note that the integral over the momentum $p$ is divergent as
it is for the ordinary Coleman-Weinberg potential in $T = \hat{\mu}_B =\hat{\mu}_F= 0$ case.
In \cite{Arai:2013ema}, we have examined the large-$p$ 
behavior of the effective potential and carefully extract the 
divergent piece. 
The effective potential consists of two parts: One is the Coleman-Weinberg potential
and the other is terms to which temperature and chemical potentials contribute.
The divergent parts reside only in the Coleman-Weinberg potential, and
so the ordinary renormalization procedure is available.

For general models, finding the analytic solutions to the equations
\eqref{eq:chi_eq} is difficult. We will therefore perform
the numerical analysis of the thermal effective potential 
to find the dependence of vacua on $T$, $\hat{\mu}_{B}$ and $\hat{\mu}_F$.
Before that, in the next section, we introduce the model \cite{Shih:2007av}
that exhibits the vacuum where supersymmetry and R-symmetry are broken at 
$T =\hat{\mu}_{B}= \hat{\mu}_F=0$.

%
%
\section{The model and meta-stable vacua at $T= \hat{\mu}_B=\hat{\mu}_F= 0$}
In this section, we introduce the generalized O'Raifeartaigh model
\cite{Shih:2007av} that
exhibits the R-symmetry breaking at supersymmetry breaking meta-stable vacua.
The generalized O'Raifeartaigh model consists of fields with R-charges
other than 0 or 2. 
The superpotential of the model is 
\begin{align}
W = f \hat{X} + \frac{1}{2} (M^{ij} + X N^{ij}) \Phi_i \Phi_j, \quad \det M
 \not= 0, \quad \det (M + \hat{X} N) \not= 0,
\label{eq:superpotential}
\end{align}
where $\hat{X}$ and $\Phi^i \ (i=1, \cdots ,N)$ are $\mathcal{N} = 1$ chiral
superfields in four dimensions. The interaction matrices $M$ and $N$ are
constant and complex, and $f$ is a real parameter. 
The component expansion of the chiral superfields in the chiral basis is
given by 
\begin{eqnarray}
\begin{aligned}
& \Phi_i = \ \phi_i + \sqrt{2} \psi_i \theta + F_i \theta^2,  \\
& \hat{X} = \ X + \sqrt{2} \psi_X \theta + F_X \theta^2, 
\end{aligned}
\end{eqnarray}
where $\theta$ is the Grassmann coordinate of the superspace.
The model has a $U(1)$ R-symmetry and admits a local
supersymmetry breaking vacuum at one-loop if
the following conditions for the R-charge are satisfied:
\begin{eqnarray}
\begin{aligned}
& R (\Phi_i) + R (\Phi_j) = 2, \quad \textrm{for } M^{ij} \not= 0, \\
& R (\Phi_i) + R (\Phi_j) = 0, \quad \textrm{for } N^{ij} \not= 0, \\
& R (X) = 2.
\end{aligned}
\end{eqnarray}
Here $R (\Phi_i)$ stands for the R-charge of the superfield $\Phi_i$.

We consider the simplest model studied in 
\cite{Shih:2007av}. 
The model consists of the three chiral superfields $\Phi_i \ (i=1,2,3)$ with
the R-charge assignment $R (\Phi_1) = -1$, $R(\Phi_2) = 1$, $R (\Phi_3)
= 3$ and a modulus chiral superfield $\hat{X}$ with $R(\hat{X}) = 2$.
The interaction matrices are given by 
\begin{align}
M = 
\left(
\begin{array}{ccc}
0 & 0 & m_1 \\
0 & m_2 & 0 \\
m_1 & 0 & 0
\end{array}
\right), \qquad 
N = 
\left(
\begin{array}{ccc}
0 & \lambda & 0 \\
\lambda & 0 & 0 \\
0 & 0 & 0
\end{array}
\right),
\label{eq:int_matrices}
\end{align}
where the parameters $m_1$, $m_2$, $\lambda$ are chosen such that they
are real.
For the superpotential \eqref{eq:superpotential} with the interaction
matrices \eqref{eq:int_matrices}, the scalar potential is calculated to
be 
\begin{align}
V = |\lambda \phi_1 \phi_2 + f|^2 
+ | \lambda X \phi_2 + m_1 \phi_3|^2 
+ | \lambda X \phi_1 + m_2 \phi_2|^2
+ | m_1 \phi_1 |^2.
\label{eq:scalar-potential}
\end{align}
Supersymmetry is spontaneously broken at the extrema of the potential:
\begin{align}
V = f^2, \quad \phi_i = 0, \quad X: \textrm{arbitrary}.
\label{eq:local_vacua}
\end{align}
Here vacua are degenerate and are parametrized by pseudo modulus field $X$.

Even though \eqref{eq:scalar-potential} has no static solution of $V=0$,
in addition to the above vacuum, 
there is a supersymmetric vacuum which lies in the following runaway direction
\begin{align}
X = 
\left(
\frac{m_1^2 m_2 \phi_3^2}{\lambda^2 f}
\right)^{\frac{1}{3}}, \qquad 
\phi_1 = 
\left(
\frac{f^2 m_2}{\lambda^2 m_1 \phi_3}
\right)^{\frac{1}{3}}, \qquad 
\phi_2= -\left({f m_1 \phi_3 \over \lambda m_2} \right)^{1/3}
\qquad 
\phi_3 \to \infty.
\end{align}
The vacuum \eqref{eq:local_vacua} is a local minimum of the potential
provided that the following bound is satisfied
\begin{align}
|X| < \frac{m_1}{\lambda} \frac{1 - y^2}{2y}, \qquad 
y = \frac{\lambda f}{m_1 m_2}. \label{region}
\end{align}
When $X$ exceeds the above bound, tachyonic modes along $\phi_i$
directions appear and the local vacuum becomes unstable.
The one-loop effective potential of the pseudo modulus $X$ at $\phi_i =
0$ is 
determined by the ordinary Coleman-Weinberg potential.
The effective potential is expanded around the origin $X=0$: 
\begin{align}
V^{(1)} = \mathrm{const.} + m^2_X |X|^2 + \frac{1}{4} \lambda_X |X|^4 +
 \mathcal{O} (|X|^6),
\end{align}
where the coefficients are given by 
\begin{eqnarray}
\begin{aligned}
m_X^2 =& \  - \frac{m_1^2 \lambda^2 y^2}{8 \pi^2} \frac{r^2 (2 r^2 (r^2 + 3)
 \log r - (3 r^4 - 2 r^2 - 1))}{(r^2 - 1)^3} + \mathcal{O} (y^4), \\
\lambda_X =& \ \frac{3 \lambda^4 y^2}{8 \pi^2} 
\frac{r^2 (12 r^2 (r^4 + 5 r^2 + 2) \log r - 19 r^6 - 9 r^4 + 27 r^2 +
 1)}{(r^2 - 1)^5} + \mathcal{O} (y^4), \\
r =& \ \frac{m_2}{m_1}.
\end{aligned}
\end{eqnarray}
The existence of the meta-stable vacuum where the
R-symmetry is broken depends on the parameters $(r,y)$, 
which are searched by numerical analysis.
Fig. \ref{fg:T0mu0} shows the allowed parameter region (black) satisfying 
$m_X^2 < 0$ and $\lambda_X > 0$ where R-symmetry is broken 
by non-zero VEV $X\not=0$.
It is also shown that the vacuum 
can be long-lived for small $y$; therefore it is meta-stable.
\cite{Shih:2007av}.
The other regions consist of two parts. 
One is a part shown in 
light grey (the left part in the plot) where a local supersymmetry breaking minimum locates 
at $X=0$. The other is a part shown in 
dark grey (the right-upper part in the plot) where
the potential monotonically decreases for (\ref{region})%
\footnote{
To draw the plot in Fig. \ref{fg:T0mu0},
we employ the following criteria for the R-symmetry breaking vacua.
We calculate the effective potential as a function of $X$
within the bound (\ref{region}),
and search the point that gives the minimum of the potential.
If the minimum is at $X=0$, that point is a local vacuum
where the R-symmetry is not broken (light gray).
If the minimum is at the nearest bound
$X \sim \frac{m_1}{\lambda} \frac{1 - y^2}{2y}$,
there is no meta-stable vacuum (dark gray).
Otherwise, that point is a local vacuum with R-symmetry breaking (black).
The plots in Fig. \ref{fig:Finite-T} is drawn using the same rule.
}.
\begin{figure}[htb]
\begin{center}
\includegraphics[bb = 0 0 300 400, scale=.6]{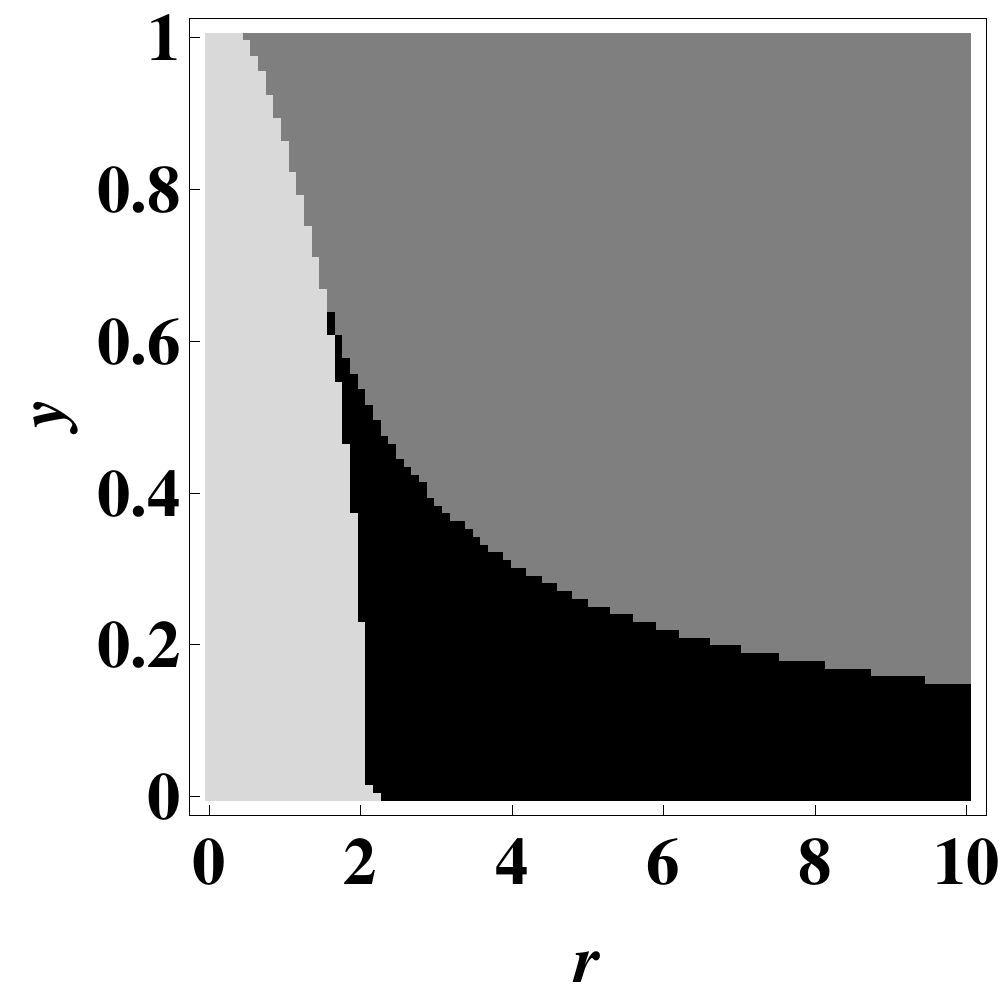}
\end{center}
\caption{The plot shows the parameter region in the $(r,y)$-plane where 
R-symmetry is broken (shown in black). The left region in the plot 
(shown in light grey) shows meta-stable vacua at $X=0$
and R-symmetry is preserved at the vacuum.
In the right-upper region (shown in dark grey) the potential monotonically
decreases for (\ref{region}).}
\label{fg:T0mu0}
\end{figure}

In the next section, we study the effects of the finite temperature and
the non-uniform chemical potentials in the generalized O'Raifeartaigh model.

%
%
\section{Meta-stable vacuum at finite temperature and chemical potentials}
Now we introduce finite temperature $T$ and chemical potentials 
in the generalized O'Raifeartaigh model.
We begin with the thermal effective potential for 
vanishing chemical potential \cite{MoSc}.
The thermal effective potential is given by
\begin{align}
V = V_{\mathrm{tree}} + V^{\mathrm{CW}}_{\mathrm{eff}} + V^{(1)}_B +
 V^{(1)}_F, 
\label{eq:effective_pot1}
\end{align}
where $V_{\mathrm{tree}}= f^2$ and $V^{\mathrm{CW}}_{\mathrm{eff}}$ is 
the Coleman-Weinberg potential of the pseudo modulus $X$,
\begin{align}
V^{\mathrm{CW}}_{\mathrm{eff}} (X) = \frac{1}{64 \pi^2} \mathrm{Tr}
\left[
m^4_B \log \frac{m^2_B}{\Lambda^2} - m^4_F \log \frac{m^2_F}{\Lambda^2}
\right].
\end{align}
Here $\Lambda$ is the dynamical cutoff scale
and $m_B^2$ and $m_F^2$ are squared mass matrices at $X$ and 
$\phi_1=\phi_2=\phi_3=0$, which are read from the superpotential 
(\ref{eq:superpotential}) as
\begin{eqnarray}
m_B^2&=&\left(
\begin{array}{cccccc}
1+X^2 & r X & 0 & 0 & ry & 0 \\
rX & r^2+X^2 & X & ry & 0 &0 \\
0 & X & 1 & 0 & 0 & 0 \\
0 & ry & 0 & 1+X^2 & rX & 0 \\
ry & 0 &0 & rX & r^2+X^2 & X \\
0 & 0 & 0 & 0 & X & 1
\end{array}
\right),
\label{eq:boson_mass}
\\
m_F^2&=&\left(
\begin{array}{cccccc}
1+X^2 & r X & 0 & 0 & 0 & 0 \\
rX & r^2+X^2 & X & 0 & 0 &0 \\
0 & X & 1 & 0 & 0 & 0 \\
0 & 0 & 0 & 1+X^2 & rX & 0 \\
0 & 0 &0 & rX & r^2+X^2 & X \\
0 & 0 & 0 & 0 & X & 1
\end{array}
\right).
\label{eq:fermion_mass}
\end{eqnarray}
Here the mass matrices are given in the real basis of $\Phi_i$:
Since there are three scalars and 3 Weyl fermions, in the real basis
each matrix has the size of $6\times 6$.
We have also divided them by $m_1^2$ after normalizing the field $X$,
\begin{eqnarray}
X\rightarrow {fX \over rym_1}.
\end{eqnarray} 
The bosonic and fermionic contributions to the thermal effective
potential are\footnote{
In \eqref{eq:boson_1-loop} and \eqref{eq:fermion_1-loop}, the degrees of
freedom of the scalar and the fermion differ by factor 2 since there are
one complex scalar and one Dirac fermion. In the present case, due to
supersymmetry, the degrees of freedom of the scalar and fermion are the
same. Therefore the factors in front of \eqref{eq:thermal_bos} and
\eqref{eq:thermal_fer} are the same.
}
\cite{Qu}
\begin{align}
& V^{(1)}_B (\phi_{\mathrm{cl}}) = \frac{T^4}{2\pi^2} 
\int^{\infty}_0 \! d x \ x^2  \mathrm{Tr}
\log \left(1 - e^{-\sqrt{x^2 + \beta^2 m^2_B}} \right), 
\label{eq:thermal_bos}
\\
& V^{(1)}_F (\phi_{\mathrm{cl}}) = - \frac{T^4}{2\pi^2} 
\int^{\infty}_0 \! d x \ x^2  \mathrm{Tr} 
\log \left(1 + e^{-\sqrt{x^2 + \beta^2 m^2_F}} \right).
\label{eq:thermal_fer}
\end{align}
Let us examine the phase structure of vacuum numerically.
The parameter regions that allow the R-symmetry breaking meta-stable vacuum
for $T=0.2, 0.5$ and $0.8$ in the $(r,y)$-plane are shown in Fig. \ref{fig:Finite-T}.
As in the case for vanishing temperature and chemical potential, 
there are three kinds of parameter regions.
The region shown in black
indicates the allowed region for the R-symmetry breaking vacua.
One observes that the allowed regions become smaller as $T$ increases.
This is expected because the effect of a finite $T$ behaves like
a positive mass for the pseudo modulus $X$ at high temperature.
Another region shown in light grey displays meta-stable 
vacua located at $X=0$, where R-symmetry is preserved.
The region shown in dark grey displays where the potential monotonically
decreases for (\ref{region}).
\begin{figure}[t]
\begin{center}
\subfigure[$T=0.20$]
{
\includegraphics[bb=0 0 300 300, scale=.5]{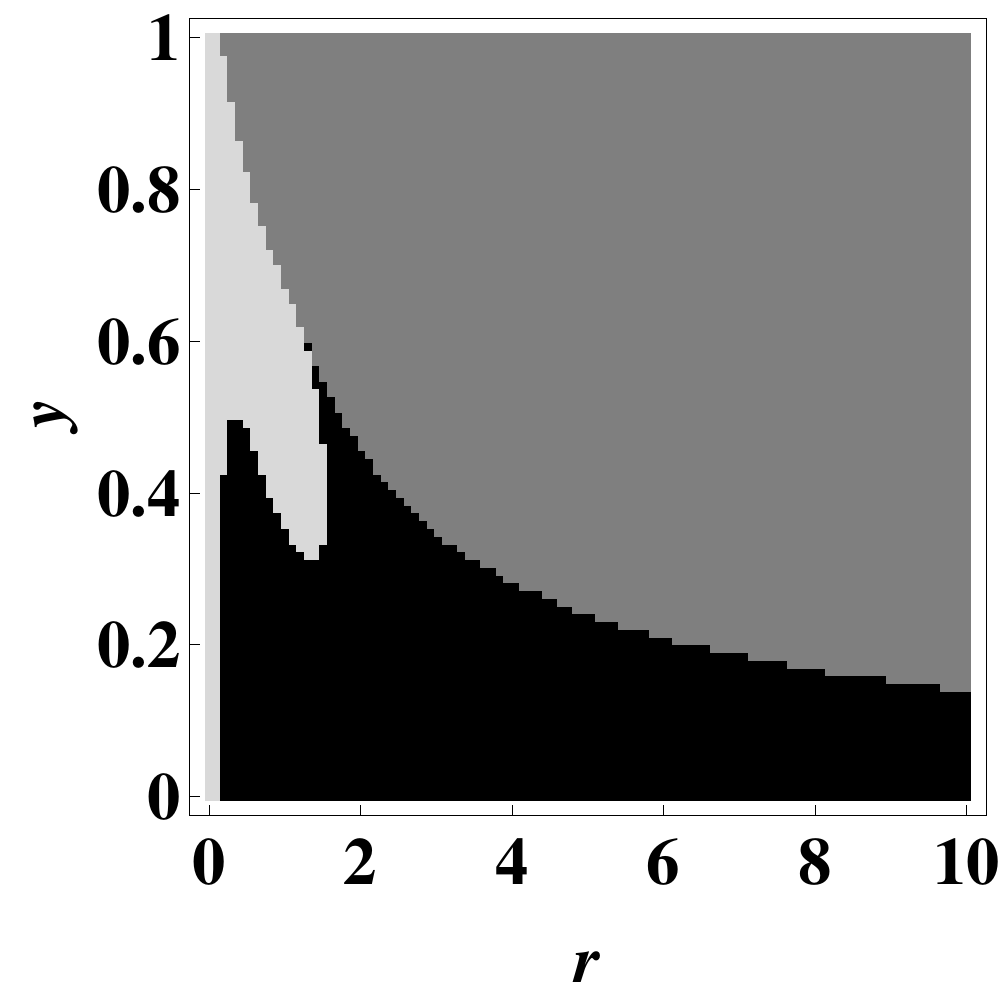}
}
\subfigure[$T=0.50$]
{
\includegraphics[bb=0 0 300 300, scale=.5]{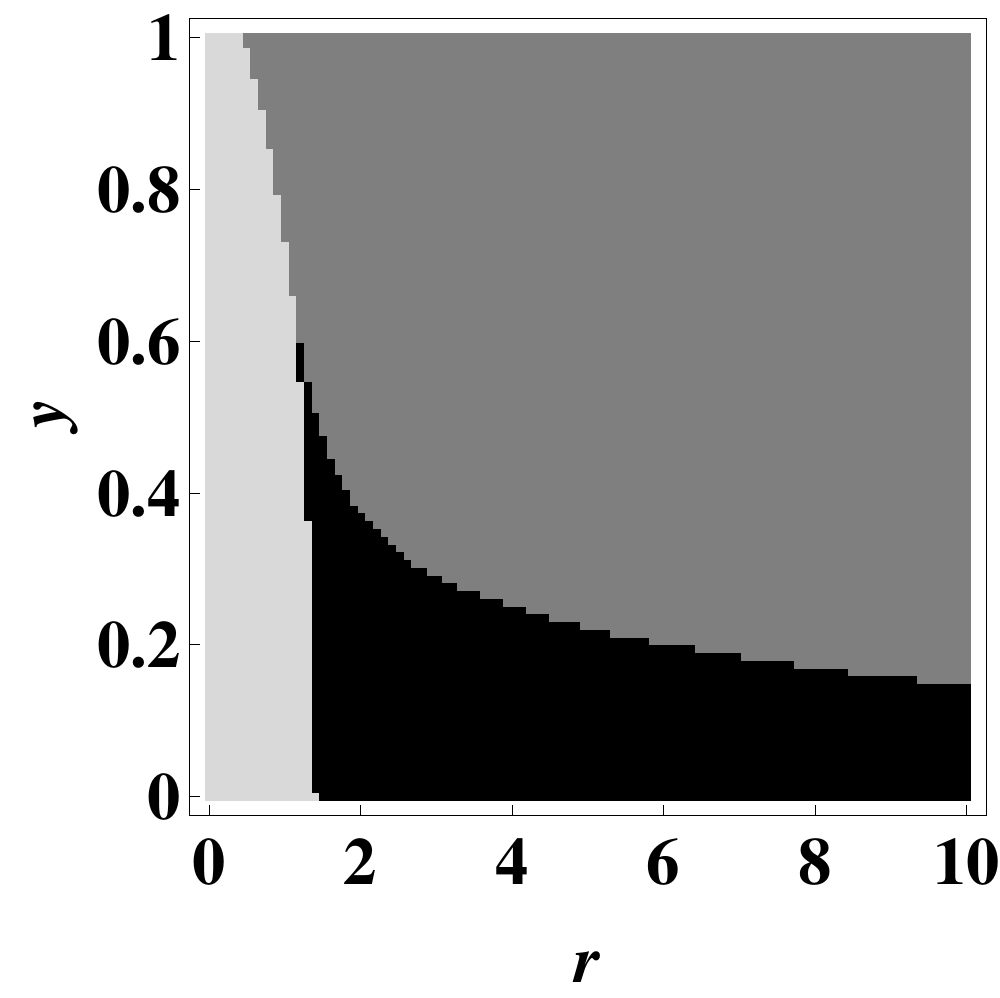}
}
\subfigure[$T=0.8$]
{
\includegraphics[bb=0 0 300 300, scale=.5]{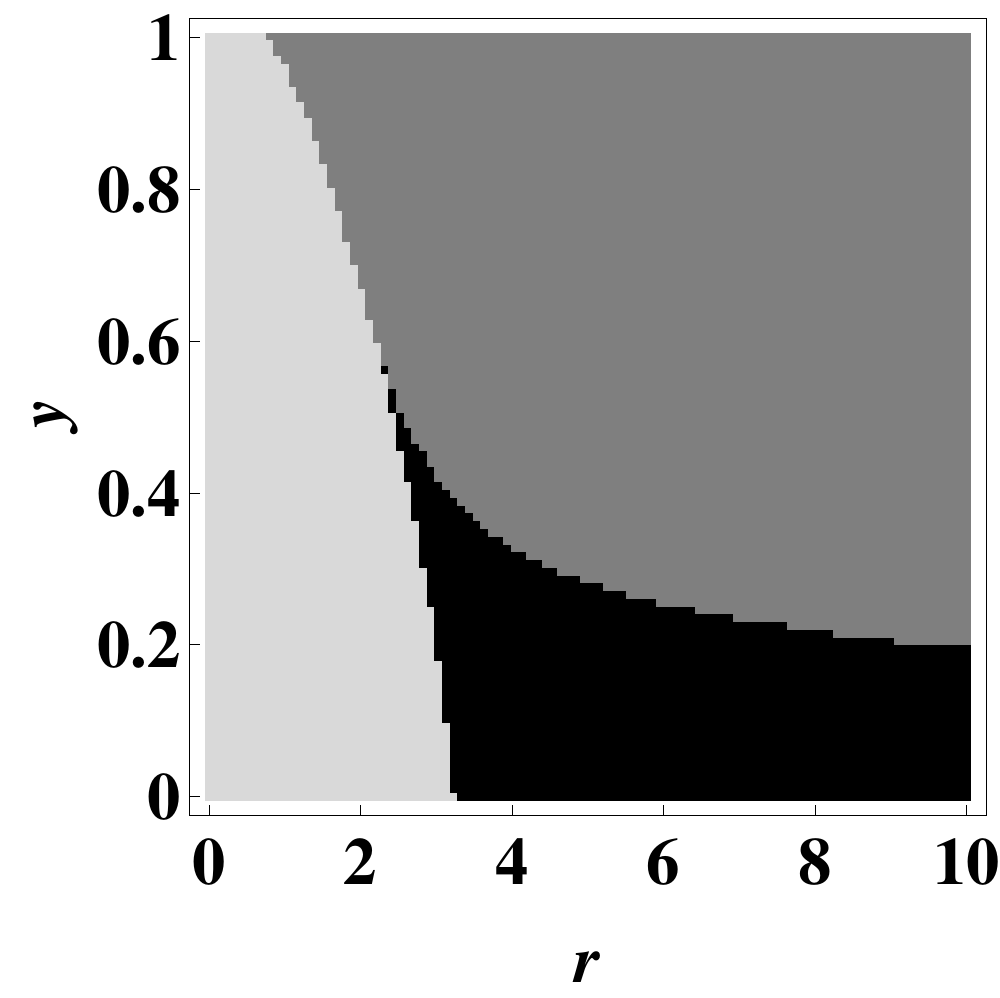}
}
\end{center}
\caption{
Plots of allowed regions (shown in black) where the R-symmetry is broken at meta-stable vacuum 
for $T=0.2$, $T=0.5$ and $T=0.8$ with $\Lambda=1$. The left region (shown in
light grey) shows the meta-stable vacuum at $X=0$
and R-symmetry is preserved at the vacuum.
In the right-upper region (shown in dark grey) the potential monotonically
decreases for (\ref{region}).
}
\label{fig:Finite-T}
\end{figure}

Now we introduce the chemical potentials. 
The thermal effective potential with the chemical potentials is given by
\begin{align}
V = V_{\mathrm{tree}} + V^{\mathrm{CW}}_{\mathrm{eff}} + V^{(1)\beta,\mu}_B +
 V^{(1)\beta,\mu}_F. 
\label{eq:effective_pot}
\end{align}
The chemical potentials in the generalized O'Raifeartaigh model are proportional to 
R-charges.
In the $\phi_i$ sector, they are of the form
%
\begin{eqnarray}
 \hat{\mu}_B=\mu {\rm diag}(-1,1,3,-1,1,3),\quad \hat{\mu}_F=\mu {\rm diag}(-2,0,2,-2,0,2).
\end{eqnarray}
In addition, 
there is the chemical potential in the $X$ sector which
provides the tachyonic mass $- 4 \mu^2 $ to the modulus field $X$.
It leads to the following tree level potential
\begin{align}
V_{\mathrm{tree}} (X) = f^2 - 4 \mu^2 X^2,
\end{align}
where we take $X$ to be real.
The one-loop parts of the thermal effective potential are given by 
\eqref{eq:pot_bos} and \eqref{eq:pot_fer}, 
while $V_{\mathrm{eff}}^{\mathrm{CW}}$ contains the Coleman-Weinberg potential
for the vanishing chemical potential and a finite correction that
depends on not temperature but only chemical potential.

We numerically investigate the potentials with several temperatures at a
specific point in the $(r,y)$-plane.
For vanishing chemical potential, as explained above there are three kinds
of parameter regions:

(1) Vacuum preserving the R-symmetry at $X=0$
(shown in light grey in Fig. \ref{fig:Finite-T}). 

(2) The potential monotonically decreases along $X$.
So the point at the origin of $\phi_i$ becomes unstable 
(shown in dark grey).

(3) Vacuum where the R-symmetry is spontaneously broken (shown in black).

Let us turn on the chemical potentials.
First we examine the case (1). Fig. \ref{fig:Finite-T-mu1} shows the effective potentials for 
$(r,y)=(1.2,0.4)$ and $T=0.2$ (left), $(r,y)=(1.3,0.2)$ and $T=0.5$ (middle), and
$(r,y)=(2.6,0.4)$ and $T=0.8$ (right).
\begin{figure}[t]
\begin{center}
\begin{tabular}{ccc}
  \begin{minipage}{5cm}
   \begin{center}
    \includegraphics[bb=0 0 300 300, scale=.35]{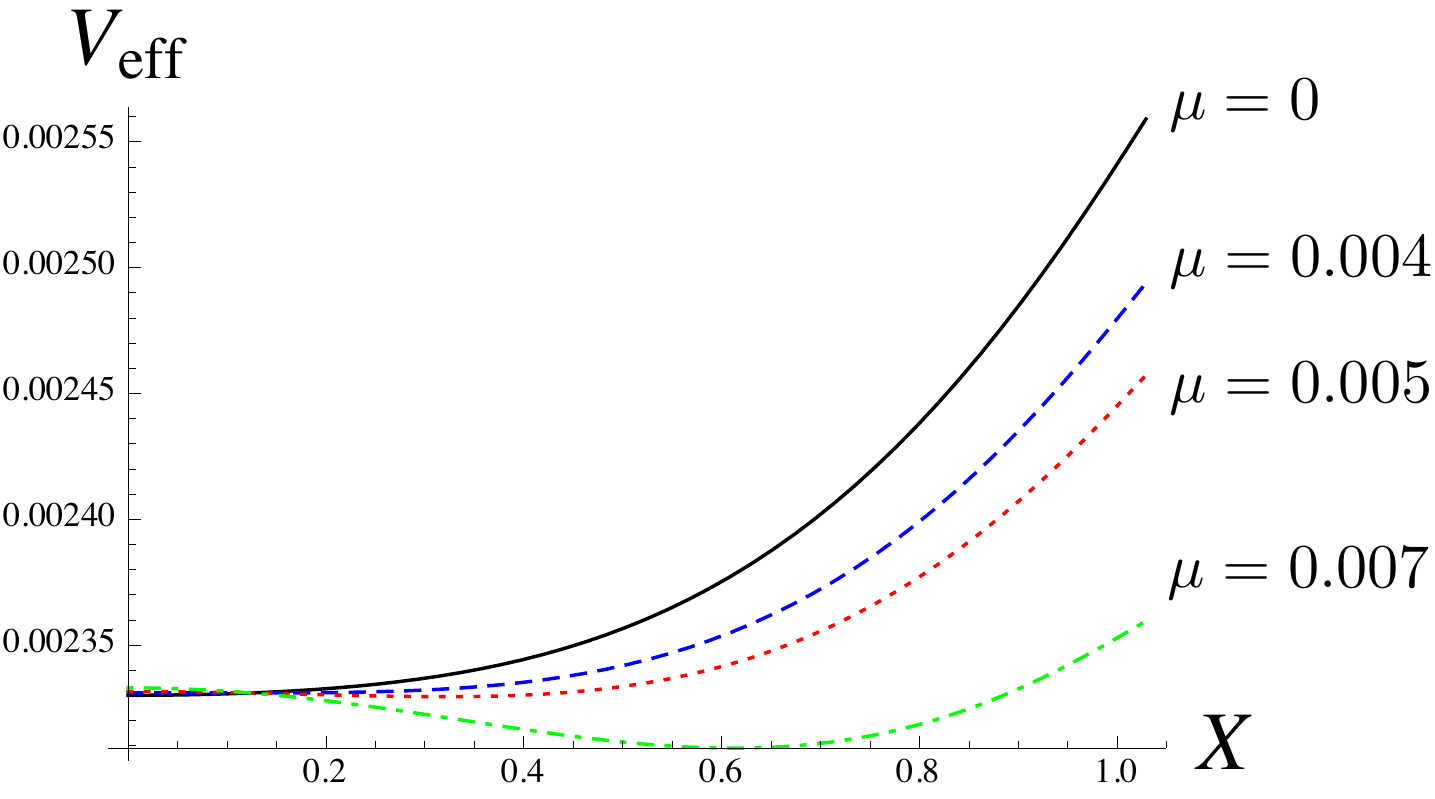}
   \end{center}
  \end{minipage}
&
  \begin{minipage}{5cm}
   \begin{center}
    \includegraphics[bb=0 0 300 300, scale=.35]{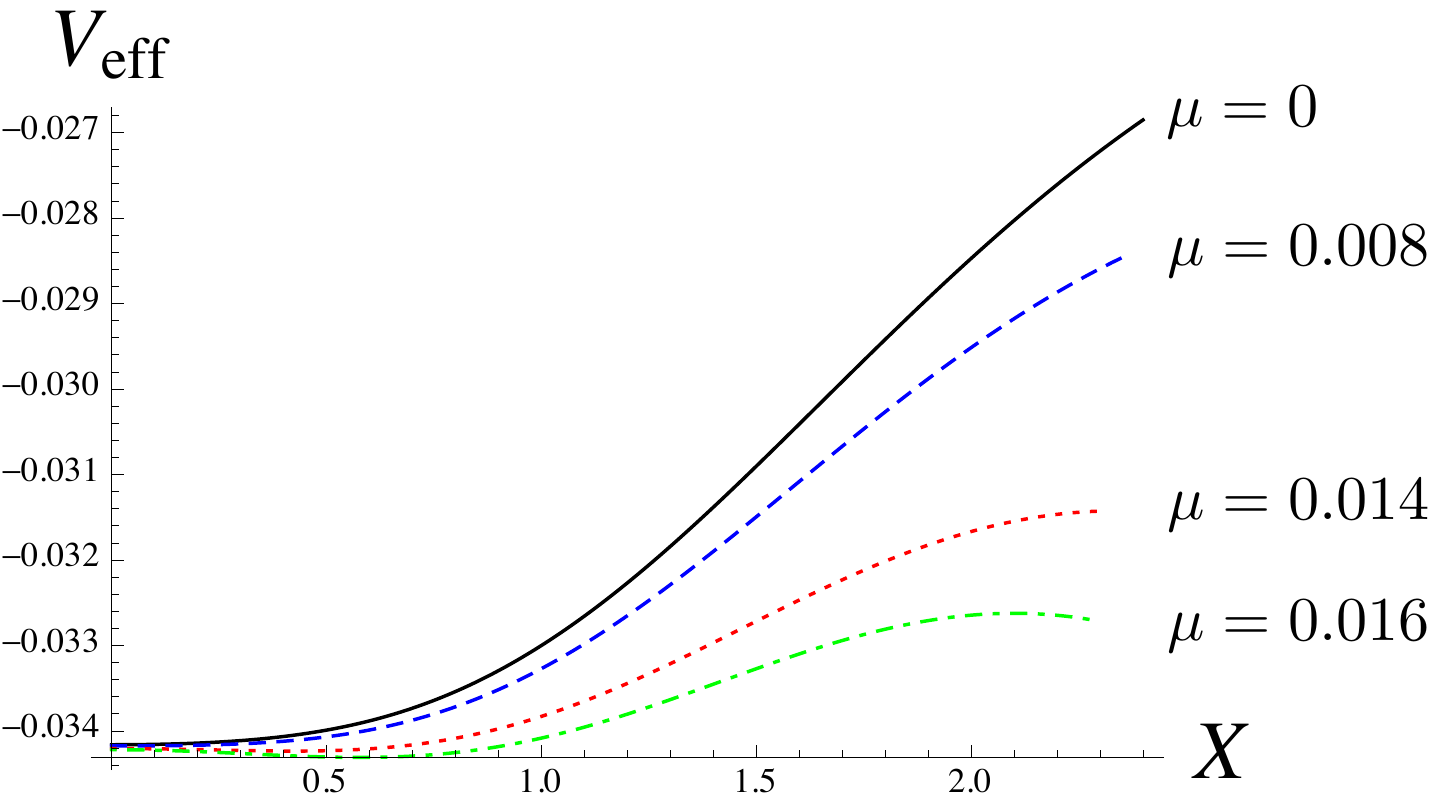}
   \end{center}
  \end{minipage}
&
  \begin{minipage}{5cm}
   \begin{center}
    \includegraphics[bb=0 0 300 300, scale=.35]{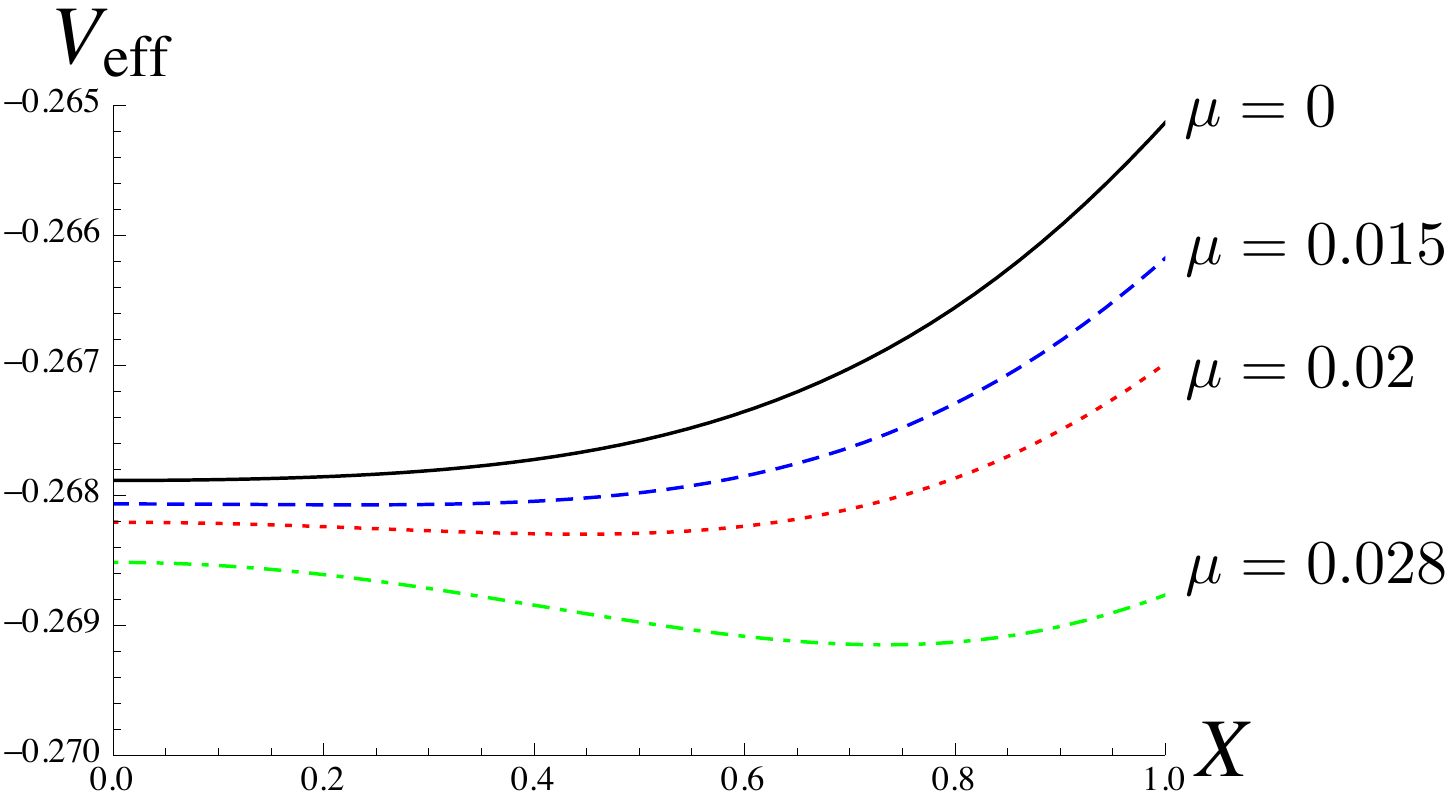}
   \end{center}
  \end{minipage}
\end{tabular}
\end{center}
\caption{
Plots of the potentials for $(r,y)=(1.2,0.4)$ and $T=0.2$ (left), $(r,y)=(1.3,0.2)$ 
and $T=0.5$ (middle), and $(r,y)=(2.6,0.4)$ and $T=0.8$ (right).
}
\label{fig:Finite-T-mu1}
\end{figure}
In this region, the origin of $X$ becomes a vacuum for $\mu = 0$ and
there are no R-symmetry breaking meta-stable vacua (solid curves).
However, when $\mu$ increases, since the non-uniform chemical potentials give 
a negative contribution to the mass of $X$, for 
some appropriate values of $\mu$ the 
quadratic and quartic terms of $X$ are balanced.
Then the origin of $X$ becomes unstable and a local minimum appears.
This phenomenon occurs not only for low temperature $T=0.2, 0.5$ but also occurs 
for high temperature $T=0.8$ (close to the dynamical cutoff scale $\Lambda=1$). 
A similar result is obtained 
in Ref. \cite{RiSe} where a supersymmetric model with a single flavor is considered,
but no spontaneous supersymmetry breaking is involved. Our result shows
that even in a model with spontaneous supersymmetry breaking the R-symmetry 
breaking occurs at high temperature.

\begin{figure}[t]
\begin{center}
\begin{tabular}{ccc}
  \begin{minipage}{5cm}
   \begin{center}
    \includegraphics[bb=0 0 300 300, scale=.35]{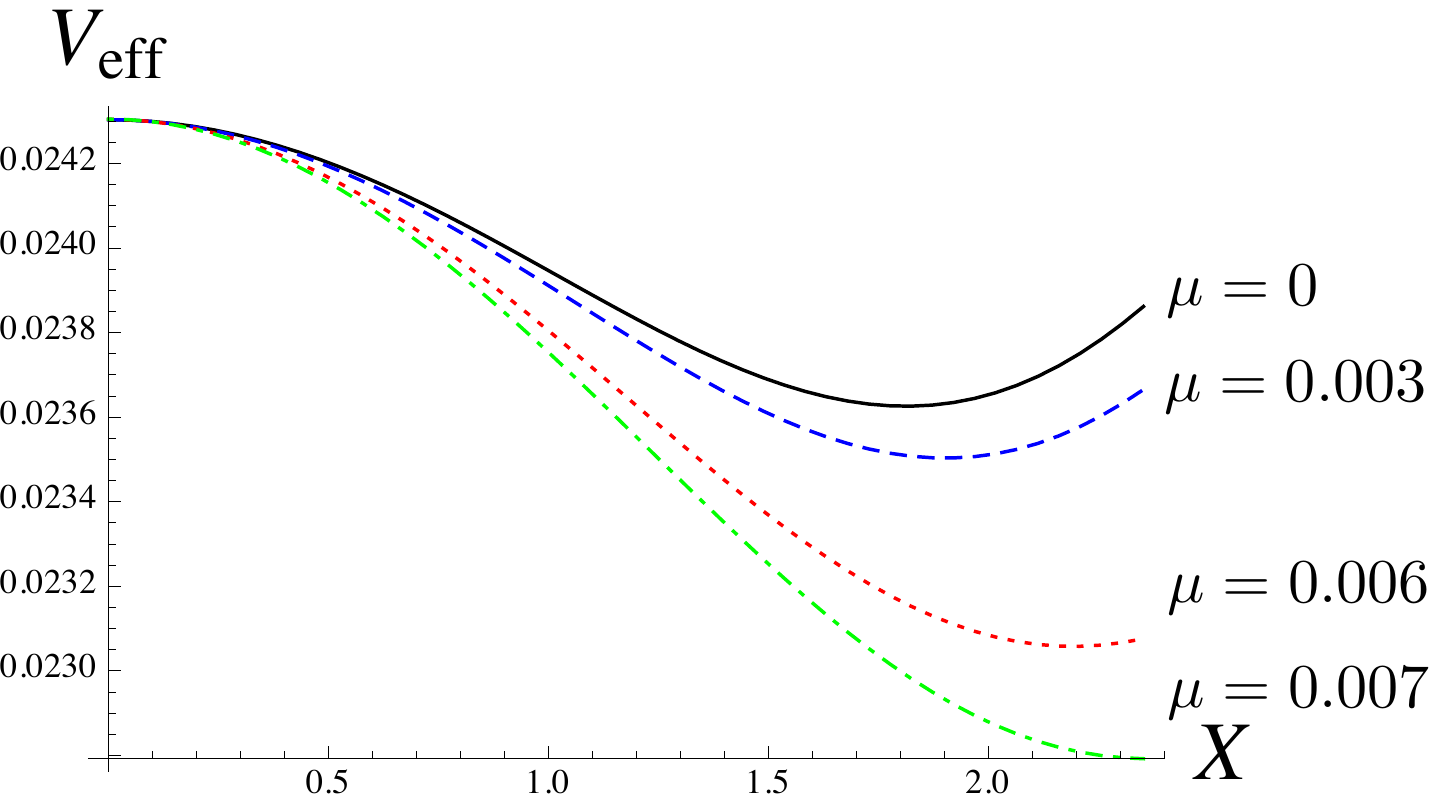}
   \end{center}
  \end{minipage}
&
  \begin{minipage}{5cm}
   \begin{center}
    \includegraphics[bb=0 0 300 300, scale=.35]{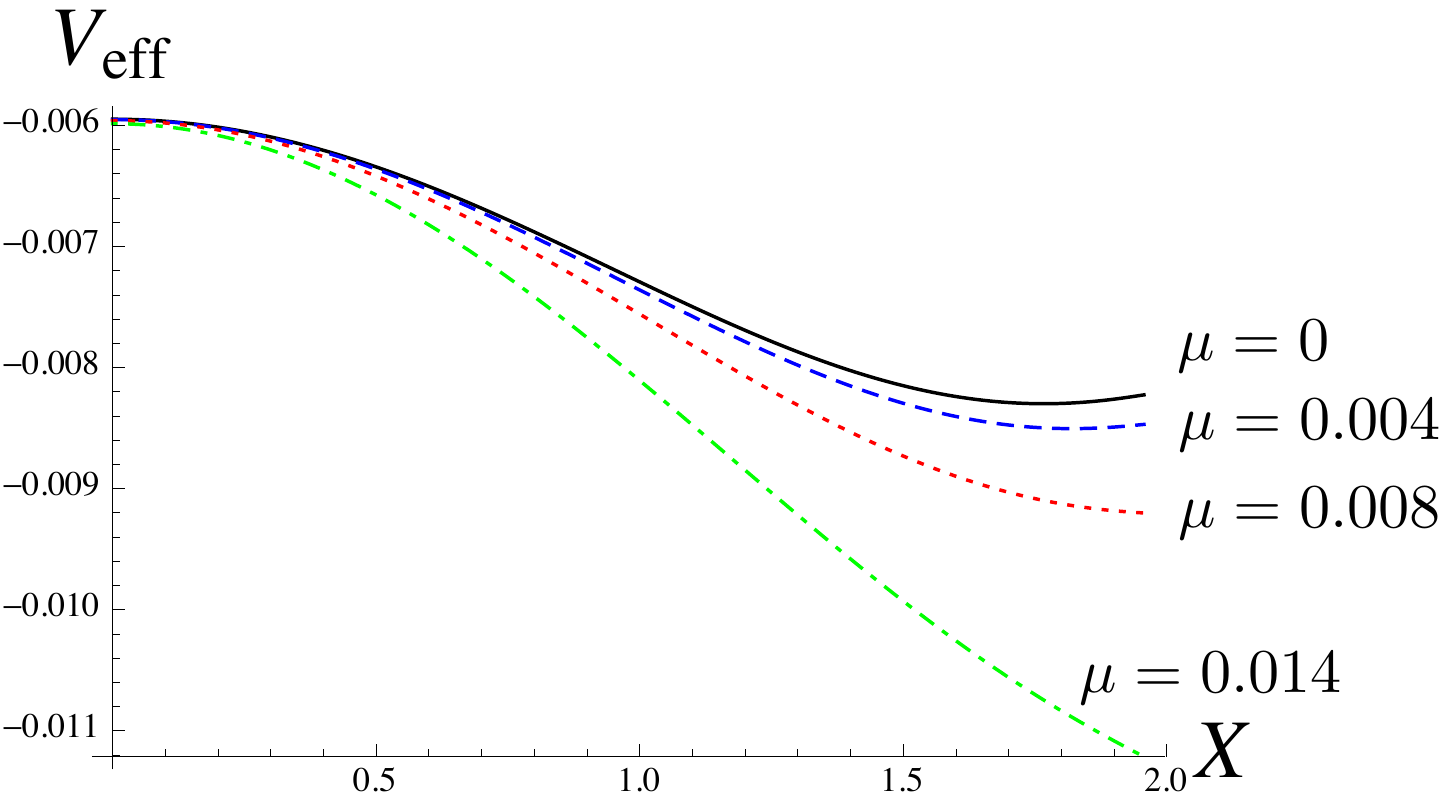}
   \end{center}
  \end{minipage}
&
  \begin{minipage}{5cm}
   \begin{center}
    \includegraphics[bb=0 0 300 300, scale=.35]{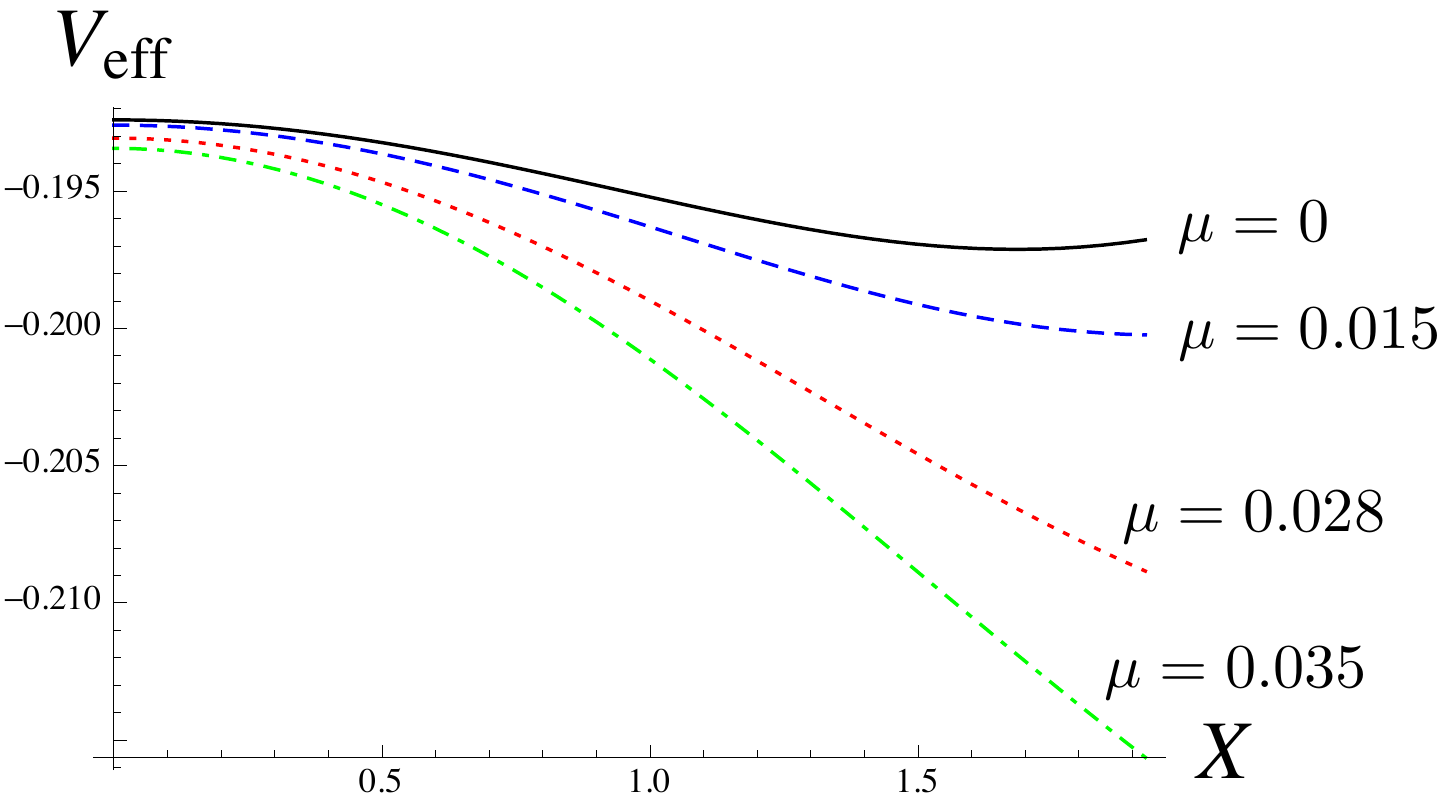}
   \end{center}
  \end{minipage}
\end{tabular}
\end{center}
\caption{
Plots of the potentials for $(r,y)=(5,0.2)$ and $T=0.2$ (left), $(r,y)=(4,0.24)$ 
and $T=0.5$ (middle), and $(r,y)=(6,0.24)$ and $T=0.8$ (right).
}
\label{fig:Finite-T-mu2}
\end{figure}

Next we consider the case (2). In this case, the potential decreases monotonically
for (\ref{region}). 
Again the non-uniform chemical potentials
contribute to the effective potential as a negative mass of $X$.
Then 
the potential keeps decreasing monotonically and therefore no new
R-symmetry breaking vacuum 
appears. 
Now let us consider the parameter region in Fig. \ref{fig:Finite-T} shown in black
near the boundary with dark grey.
As such parameters we consider
$(r,y)=(5,0.2)$ for $T=0.2$, $(r,y)=(4,0.24)$ for $T=0.5$, and $(r,y)=(6,0.24)$ for 
$T=0.8$.
The potentials for vanishing chemical potential for $T=0.2, 0.5$ and $0.8$ (solid curves)
are shown in Fig. \ref{fig:Finite-T-mu2}. 
As $\mu$ increases, the chemical potentials give rise to a negative mass of $X$
and for large chemical potentials a meta-stable vacuum is destabilized (dotted, dashed
and dot-dashed curves in Fig. \ref{fig:Finite-T-mu2}). 
This result tells us that the parameter regions of (2) become larger and
the region allowing the R-symmetry breaking reduces.

Finally, we consider the case (3), especially the
points deep inside the allowed regions. 
For instance, we take $(r,y)=(6,0.01)$ which is in a middle-bottom region
shown in black in Fig. \ref{fig:Finite-T}.
Fig. \ref{fig:Finite-T-mu3} shows the plots of the effective potentials
for $T=0.2$ (left), $0.5$ (middle) and $0.8$ (right).
For vanishing chemical potential (solid curves), there exists a meta-stable vacuum.
As like in the parameter region close to one of the case (2), 
the meta-stable vacuum for $\mu=0$ is destabilized as $\mu$ increases (dotted, dashed
and dot-dashed curves).

Considering all the results mentioned above, we conclude that while 
the allowed parameter region for $\mu=0$ to break the R-symmetry reduces,
the new parameter region to break the R-symmetry appears as $\mu$ increases.

\begin{figure}[t]
\begin{center}
\begin{tabular}{ccc}
  \begin{minipage}{5cm}
   \begin{center}
    \includegraphics[bb=0 0 300 300, scale=.35]{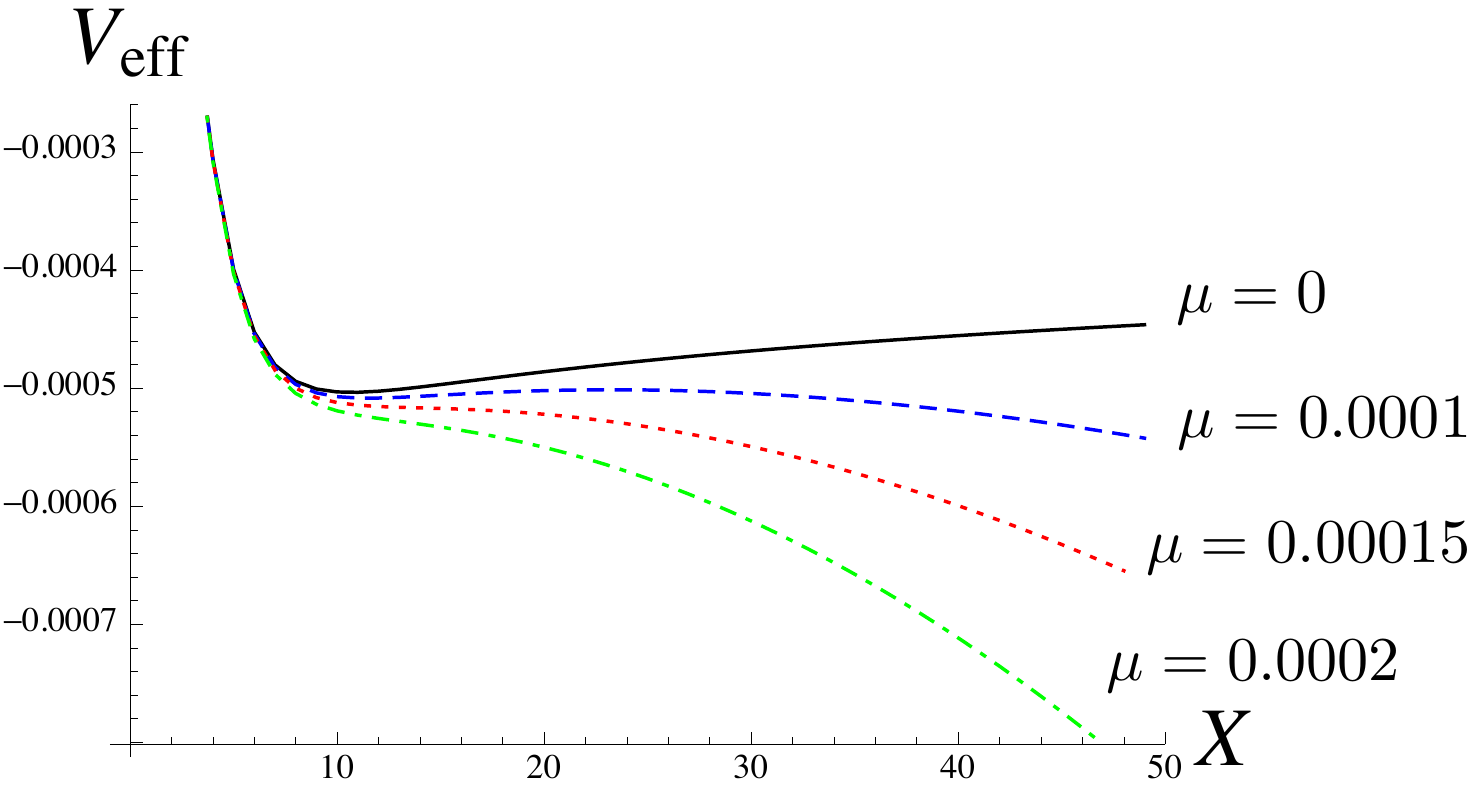}\\
   \end{center}
  \end{minipage}
&
  \begin{minipage}{5cm}
   \begin{center}
    \includegraphics[bb=0 0 300 300, scale=.35]{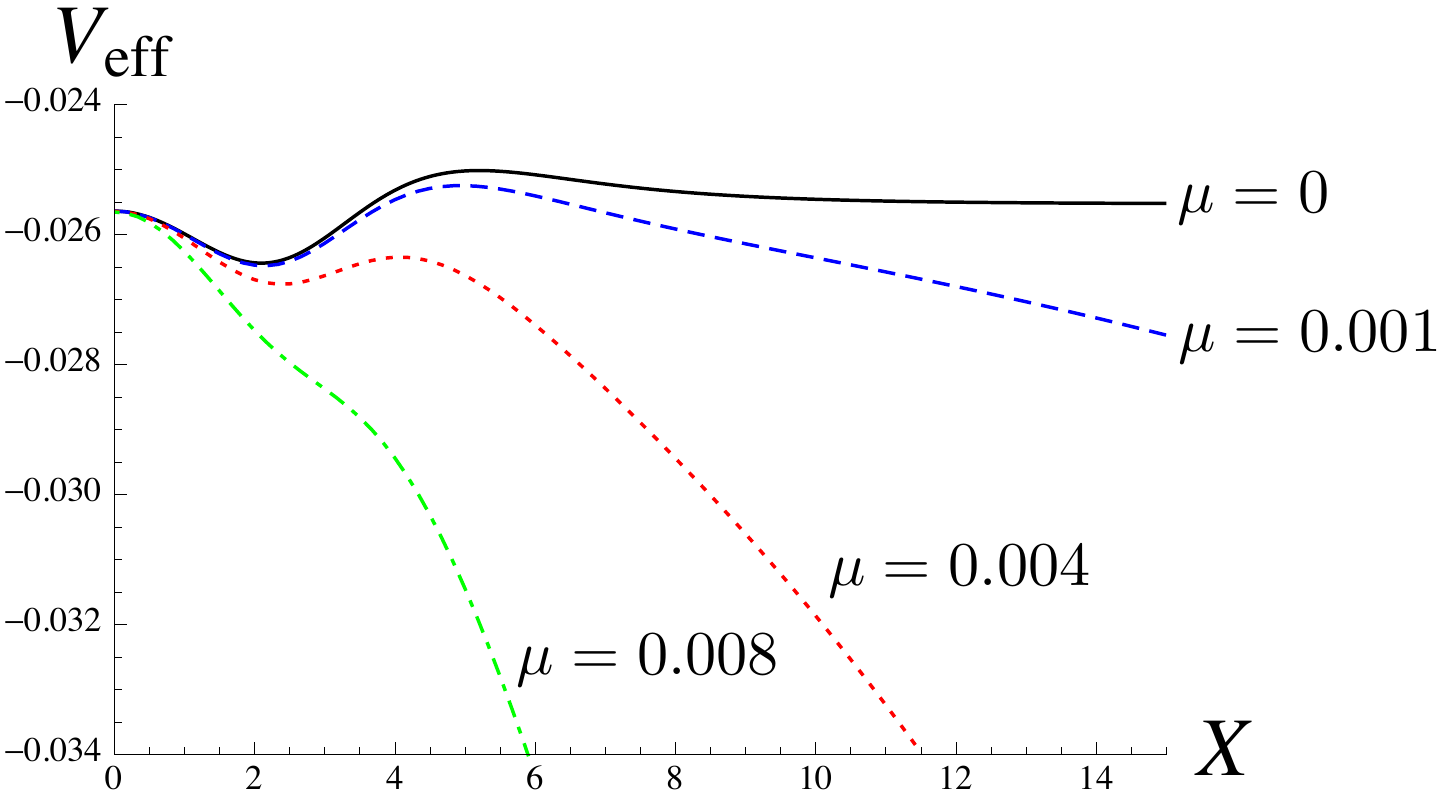}\\
   \end{center}
  \end{minipage}
&
  \begin{minipage}{5cm}
   \begin{center}
    \includegraphics[bb=0 0 300 300, scale=.35]{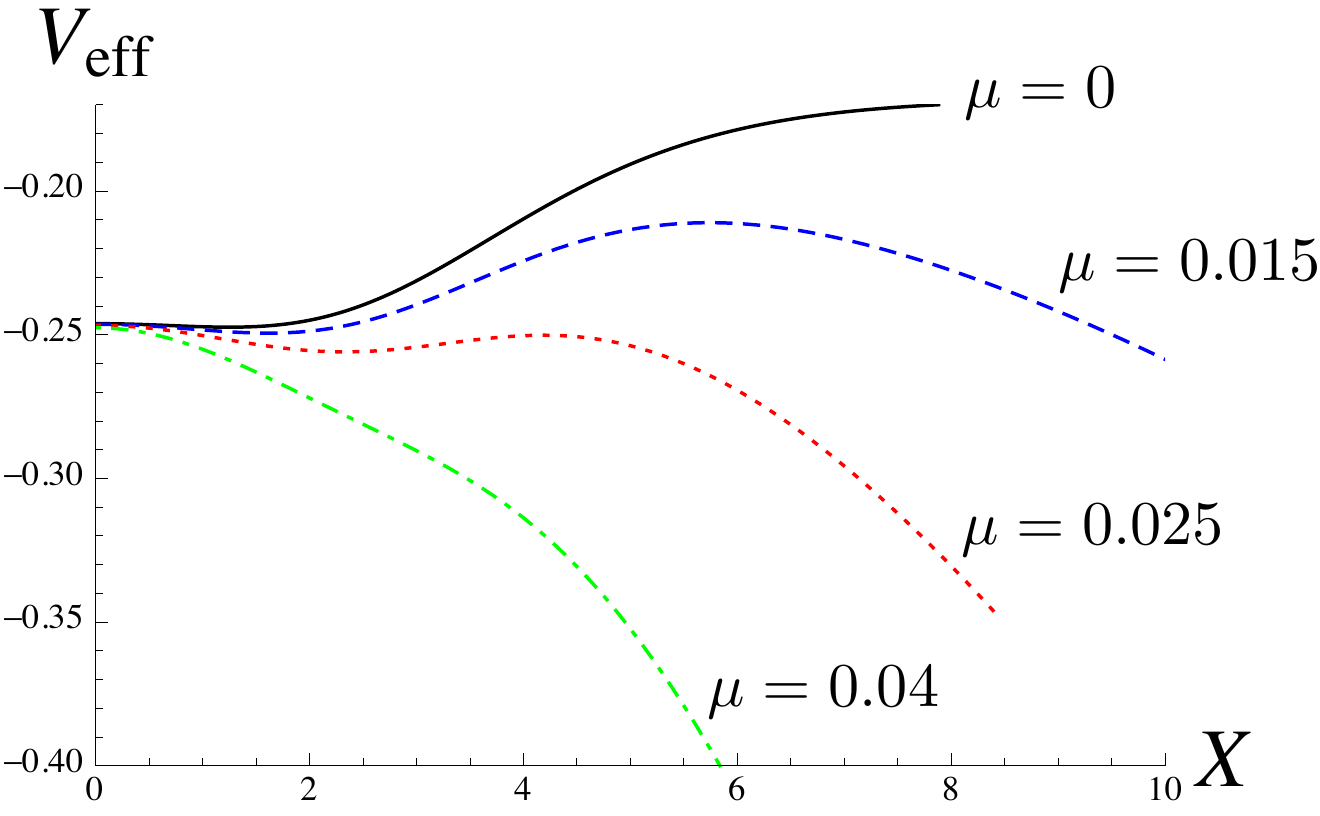}\\
   \end{center}
  \end{minipage}
\end{tabular}
\end{center}
\caption{
Plots of the potentials for $T=0.2$ (left), $T=0.5$ (middle) and $T=0.8$ (right)
with $(r,y)=(6,0.01)$. 
}
\label{fig:Finite-T-mu3}
\end{figure}

%
%
\section{Conclusion and discussions}
In this paper we have studied the meta-stable vacuum in the generalized
O'Raifeartaigh model with the non-uniform chemical potentials at finite
temperature. 
We consider the model studied in \cite{Shih:2007av} which consists of
three chiral superfields $\Phi_i \ (i=1,2,3)$ and a modulus superfield $\hat{X}$.
At zero temperature with the vanishing chemical potential, 
supersymmetry is broken at the origin of $\phi_i$ and the flat
direction is parametrized by $X$. 
The one-loop Coleman-Weinberg potential reveals that the origin of $X$
is unstable.
For the appropriate parameters $(r,y)$, 
there is a meta-stable vacuum where the VEV of $X$ is nonzero and the
R-symmetry is spontaneously broken.
The allowed parameter region is shown in Fig. 1.

When finite temperature is introduced, the temperature behaves as a
positive mass of the pseudo modulus field $X$ at the origin $\phi_i = 0$.
At high temperature, the effective mass of $X$ is large enough and the
vacuum settles down to the origin $X = 0$ and 
the R-symmetry is restored. 
This means that the allowed parameter region becomes smaller
when the temperature increases. 

However, the situation changes drastically when we introduce the
non-uniform chemical potentials. 
We have explicitly performed the numerical analysis of the thermal effective
potential with non-uniform chemical potential. 
In an appropriate parameter region, 
the chemical potential behaves as a tachyonic mass of the field $X$.
In the region shown in light grey in Fig. \ref{fig:Finite-T},
for appropriate values of $T$ and $\mu$ , we have found that 
there appears a new parameter region allowing a meta-stable
vacuum where the R-symmetry is spontaneously broken.
Since this parameter region is not inside the allowed region for $\mu =0$, 
the new vacuum is developed by the non-uniform chemical potentials.
On the other hand, we have found that the meta-stable vacua 
in a certain allowed region shown in black in Fig. \ref{fig:Finite-T}
become unstable and disappear when $\mu$ increases at fixed temperature.
Consequently, the allowed region becomes smaller in comparison with
that for the vanishing chemical.
We stress that although the allowed region of the R-symmetry breaking
vacua is totally reduced at high temperatures and large chemical
potentials, the new parameter region allowing the R-symmetry breaking
vacua appears at finite $T$ and $\mu$.
This result opens the possibility of supersymmetry breaking meta-stable
vacua in the early stage of the Universe where the R-symmetry is broken.
It is also interesting to explore a more realistic model where meta-stable
supersymmetry and R-symmetry breaking vacua appear even at high
temperature and non-zero chemical potentials. 

As for application to a realistic model building, we briefly consider
phenomenological aspects of our model. 
In our model, effects of temperature and chemical potentials are taken into
account since as mentioned in the introduction, we consider the supersymmetry breaking
in the early Universe. For instance, we suppose that the supersymmetry breaking occurs 
after the reheating. It is possible because, as we showed, the supersymmetry can be
broken even at high temperature due to the chemical potential.
Let us discuss how the size of the chemical potential is evaluated.
In order to do that, we consider our model as a hidden sector in a gauge 
mediated supersymmetry breaking scenario \cite{GM}.
In this scenario, we introduce the messenger fields coupling to the modulus 
superfield $\hat{X}$ through the following superpotential. 
\begin{eqnarray}
 W=\hat{X}\tilde{\Phi}\Phi,
\end{eqnarray}
where $\Phi$ and $\tilde{\Phi}$ are messenger fields charged under 
the standard model gauge groups. 
In our model, since supersymmetry is broken with a certain parameter choice, 
the scalar component $X$ and the F-component $F_X$ of $\hat{X}$ develop the VEVs. 
They lead to the soft supersymmetry breaking masses through the quantum corrections:
\begin{eqnarray}
M_{\rm soft}\sim {\alpha_{\rm SM} \over 4\pi}{\langle F_X \rangle \over \langle X \rangle},
\end{eqnarray}
where $\alpha_{\rm SM}$ stands for the standard model gauge coupling.
The VEVs of $\langle X \rangle$ and $\langle F_X \rangle$ are written by the chemical potential
$\mu$, the temperature $T$, $m_1$, $m_2$ and $\lambda$.
The soft mass is phenomenologically favored to set $M_{\rm soft}\sim 1$
TeV the value of which constraints the above parameters such as the chemical potential.
In this way, it is certainly interesting to see whether our hidden sector is phenomenologically viable.

\subsection*{Acknowledgments}
The work of M. A. is supported by Grant-in-Aid for 
Scientific Research from the Ministry of Education, Culture, 
Sports, Science and Technology, Japan (No.25400280).
The work of S.~S. is supported in part by Kitasato University
Research Grant for Young Researchers.




\end{document}